\documentclass[aps,prb,superscriptaddress,twocolumn]{revtex4}
\usepackage{amssymb}
\usepackage{graphicx}
\usepackage{amsmath}
\usepackage[colorlinks=true,linkcolor=blue,citecolor=blue,urlcolor=blue]{hyperref}

\newcommand{\be}{\begin{equation}}
\newcommand{\ee}{\end{equation}}

\newcommand{\bs}{\begin{split}}
\newcommand{\es}{\end{split}}

\usepackage[normalem]{ulem}
\usepackage{amsmath,amssymb}
\usepackage{multirow}
\usepackage{xcolor,soul}
\usepackage{srcltx}
\usepackage{hyperref,graphicx}

\begin{document}

\title{Dephasing-induced mobility edges in quasicrystals}
\author{Stefano Longhi}
\thanks{stefano.longhi@polimi.it}
\affiliation{Dipartimento di Fisica, Politecnico di Milano, Piazza L. da Vinci 32, I-20133 Milano, Italy}
\affiliation{IFISC (UIB-CSIC), Instituto de Fisica Interdisciplinar y Sistemas Complejos, E-07122 Palma de Mallorca, Spain}

\begin{abstract}
Mobility edges (ME), separating Anderson-localized states from extended
states, are known to arise in the single-particle energy spectrum of certain one-dimensional lattices with aperiodic order. Dephasing and decoherence effects are widely acknowledged to spoil Anderson localization and to enhance transport, suggesting that ME and localization are unlikely to be observable in the presence of dephasing. Here it is shown that, contrary to such a wisdom, ME can be created by pure dephasing effects in quasicrystals in which all states are delocalized under coherent dynamics. Since the lifetimes of localized states induced by dephasing effects can be extremely long, rather counter-intuitively decoherence can enhance localization of excitation in the lattice. The results are illustrated by considering photonic quantum walks in synthetic mesh lattices.  
 \end{abstract}

\maketitle

{\it Introduction.}
Anderson localization and mobility edges \cite{R1,R2,R2b,R3} are fundamental concepts in the physics of disordered systems. They concern the localization behavior of quantum or classical waves in systems with disorder, and play a crucial role in different areas of physics, ranging from condensed matter physics \cite{R1,R2,R2b,R3} to ultracold atoms \cite{R4,R5,R6,R7,R8,R9,R10,R11} and disordered photonics \cite{R12,R13,R14,R15,R16,R16b,R16c}.  Mobility edges (ME)  generally refer to points or thresholds in the energy spectrum where the localization features of the wave functions change, from being exponentially localized to being spatially extended \cite{R2,R3}. The ME leads to various fundamental phenomena, such as metal-insulator transition by
varying the particle number density or disorder strength \cite{R3}, and  can
survive under perturbations and interactions \cite{R17,R18}. While in non-interacting low-dimensional systems ME are prevented and the wave functions are localized with arbitrarily small
disorder strength \cite{R2b,R19}, it is well known that ME can exist in certain one-dimensional (1D) systems with quasiperiodic order, i.e. in quasicrystals \cite{R20,R21,R22,R23,R24,R25,R26,R27,R28,R29,R30,R31,R32,R33,R34,R35,R36,R37,R37b,R37c}.\\
Anderson localization and ME are generally observable in systems with non-fluctuating disorder. Besides, decoherence and dephasing effects are known to spoil localization and to enhance transport \cite{R38,R39,R40,R41,R42,R43,R44,R45,R46,R47,R47b,R48,R49}. This evidence would suggest that dephasing effects are detrimental to enhance localization or to create ME, especially when the system does not display Anderson localization under coherent evolution.\\
In this Letter it is shown that, contrary to such a wisdom, ME can be created and localization can be enhanced by dephasing effects, even in systems which do not display Anderson localization under coherent dynamics. This unexpected result is illustrated by considering the off-diagonal Aubry-Andr\'e model \cite{R50,R51,R52,R53,R54,R55,R56}  in the delocalized phase, where dephasing effects can create ME and slow down delocalization in the lattice.  A photonic quantum walk setup in a synthetic quasicrystal is suggested as an experimentally accessible platform for the observation of dephasing-induced ME.\\
\\
{\it Model.}  We consider a tight-binding one-dimensional lattice with aperiodic order described by the Hamiltonian 
\begin{equation}
\hat{H}= -\sum_n \left( J_n \hat{a}^{\dag}_{n+1} \hat{a}_{n}+\rm{H.c.} \right)+\sum_n V_n \hat{a}^{\dag}_{n} \hat{a}_n
\end{equation}
where $J_n$ is the hopping amplitude between adjacent sites $n$ and $(n+1)$, $V_n$ is the on-site potential, and $\hat{a}^{\dag}_n$, $\hat{a}_n$ are the creation and annihilation operators of bosonic particles at lattice site $n$, satisfying the usual commutation relations. Aperiodic order can be introduced in either or both the hopping amplitudes or on-site potential.  We will focus our attention to two paradigmatic models, namely (i) the generalized diagonal Aubry-Andr\'e model \cite {R24}, corresponding to $J_n=J$,  $V_n=2A \cos (2 \pi \alpha n+\theta) / [1-B \cos (2 \pi \alpha n+\theta)]$,
with $J,A>0$, $ 0 \leq B<1$ and $\alpha$ irrational Diophantine (this model reduces to the ordinary Aubry-Andr\'e model \cite{R50} for $B=0$); and (ii) the off-diagonal Aubry-Andr\'e model \cite{R51,R52,R53,R54,R55}, corresponding to  $J_n=A+B \cos (2 \pi \alpha n+\theta)$, $V_n=0$,
with $A,B>0$. The first model displays ME in certain regions of parameter space at the energy $E_m=(2/B)(J-A)$ \cite{R24} [Fig.1(a)], whereas the second model
does not display ME and for $B<A$ all single-particle eigenstates of $H$ are extended [Fig.1(c)]. In such regimes an initial excitation in the lattice spreads ballistically, as illustrated in Figs.1(b) and (d).  In the numerical analysis we assume  $\alpha=( \sqrt{5}-1)/2$ (the golden mean) and take a lattice of finite (but large) size $L$ with periodic boundary conditions, where $L$ is a Fibonacci number and the golden mean is approximated by a rational. The localization properties of the eigenstates $\psi_n^{(l)}$ of $H$ are captured {by  the inverse participation ratio (IPR) and fractal dimension. For a normalized wave function they are defined as \cite{R3,Thouless}  $\text{IPR}_l =\sum_{n=1}^{L} | \psi^{(l)}_{n} |^{4}$ and \cite{R3,fract1,fract2,fract3}  $\beta_{l}= \lim_{L \rightarrow \infty} (  \ln {\rm IPR}_{l} ) / \ln (1/L) $.  For a localized wave function  $\beta_{l}=0$, for an extended (ergodic) wave function $\beta_{l}=1$, whereas for a critical wave function  $0<\beta_{l}<1$.} The spreading dynamics of initial single-site excitation of the lattice is captured by the time behavior of the second moment $\sigma^2(t)= \sum_n n^2 | \psi_n(t)|^2$, where $\psi_n(t)$ are the amplitude probabilities at various lattice sites at time $t$.  Dynamical delocalization corresponds to a secular growth of $\sigma^2(t)$ in time, with $\sigma^2(t) \sim t^2$ for ballistic spreading and $\sigma^2(t) \sim t$ for diffusive spreading \cite{spreading}.

Dephasing effects can be modeled using stochastic Schr\"odinger equations or quantum master equations in the Lindblad form (see e.g. \cite{
R39, R41,R43,R45,R46,R47,R48,R49,R57,R58}). The Lindblad master equation for the density matrix $\hat{\rho}$ describing dephasing effects reads \cite{R46,R49,R57,R58}
\begin{equation}
\frac{d \hat{\rho}}{dt} = -i [ \hat{H}, \hat{\rho} ] + \gamma \sum_{n=1}^{L} \left( \hat{L}_n \hat{\rho} \hat{L}_n^{\dag}-\frac{1}{2} \left\{ \hat{L}_{n}^{\dag} \hat{L}_n, \hat{\rho} \right\}  \right)
\end{equation}
 where $\hat{L}_n = \hat{a}^{\dag}_n \hat{a}_n$ is the dissipator describing pure dephasing at lattice site $n$ and $\gamma>0$ is the dephasing rate. This model conserves the number of particles. In the single-particle sector the Hilbert space,  spanned by the set of states  $|n \rangle = \hat{a}_n^{\dag} | vac \rangle$, the evolution equations for the density matrix elements $\rho_{n,m}(t)= \langle n | \hat{\rho} |m \rangle$ read \cite{R57,R58}
 \begin{eqnarray}
 \frac{d \rho_{n,m}}{dt} & = & i (J_{n-1} \rho_{n-1,m}+J_n \rho_{n+1,m}) \nonumber  \\
 & - & i (J_m \rho_{n,m+1}+J_{m-1} \rho_{n,m-1}) \\
  & + &i (V_m-V_n) \rho_{n,m}-\gamma (1- \delta_{n,m}) \rho_{n,m}. \nonumber
 \end{eqnarray}

\begin{figure}[t]
    \centering
    \includegraphics[width=0.5\textwidth]{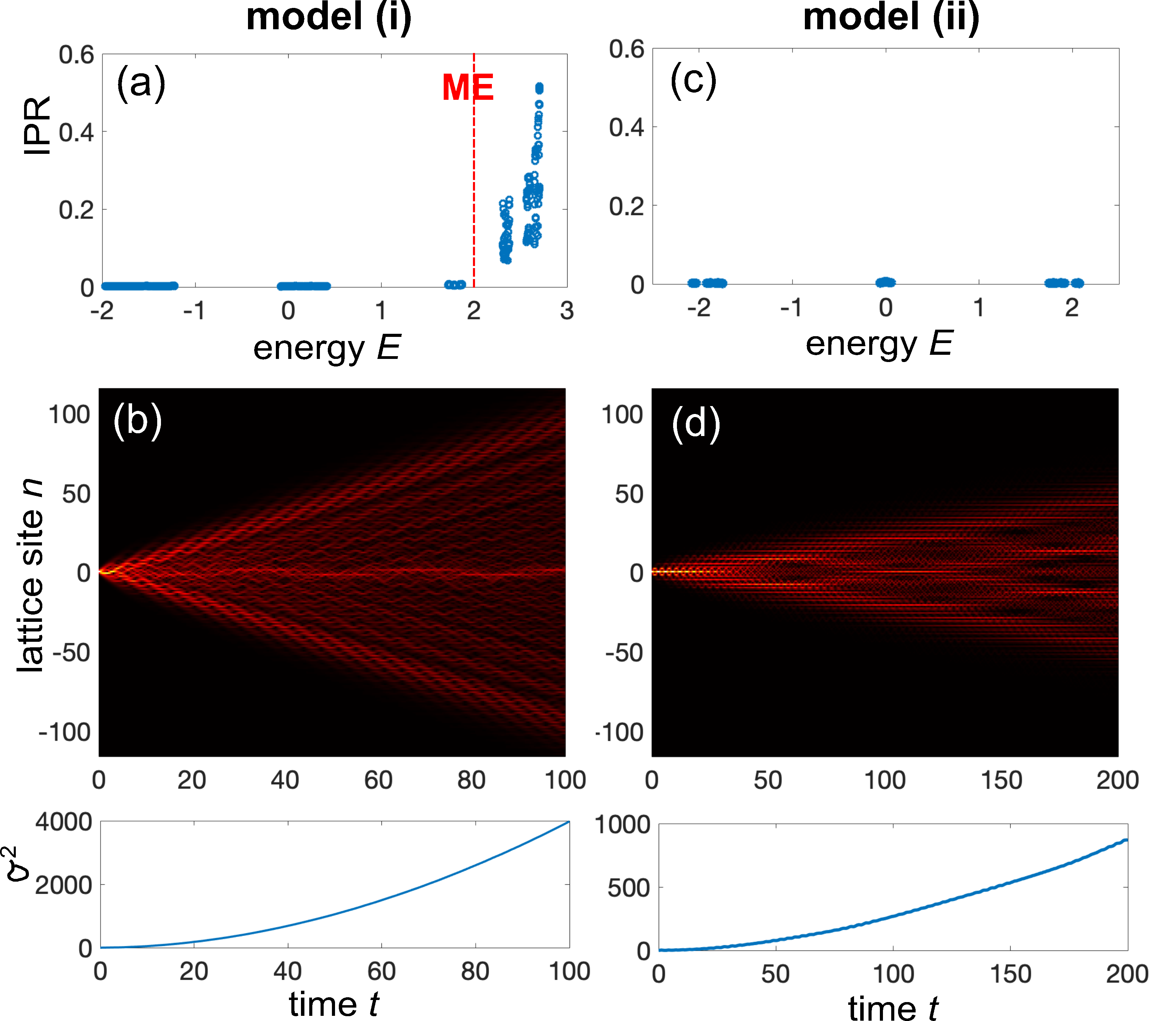}
   \caption{(a) IPR of the eigenstates of the Hamiltonian $H$ for the generalized diagonal Aubry-Andr\'e model [model (i)] for parameter values $\alpha=(\sqrt{5}-1)/2$, $J=1$, $A=0.6$, and $B=0.4$. Note the existence of a ME at the energy $E_m=2$, with extended states for $E<E_m$ and localized states for $E>E_m$. (b) Numerically computed wave spreading dynamics in the lattice (snapshot of $|\psi_n(t)|$ on a pseudocolor map) for initial excitation of site $n=0$. The lower panel in (b) depicts the corresponding behavior of the second moment $\sigma^2(t)$. (c,d) Same as (a,b), but for the off-diagonal Aubry-Andr\'e model [model (ii)] with parameter values $A=1$, $B=0.9$. Note that in this case there are not ME and all wave functions are extended. In both models the spreading in the lattice is ballistic. Lattice size is $L=987$.}
    \label{fig1}
\end{figure}

 
{\it Dephasing-induced mobility edges and localization.} 
The main question is whether dephasing effects can create ME and induce some kind of localization. To answer this question, we focus our analysis to the strong dephasing regime ($\gamma \gg |J_n|, |V_n| $), such that the coherences $\rho_{n,m}$ ($ n \neq m$) are small and the particle dynamics can be described by a classical one-dimensional hopping model \cite{R46,R47,R47b}. Basically, the time evolution of the populations $P_{n}(t)=\rho_{n,n}(t)$  is given by a classical master equation, where the quantum jumps generate the transition rates between classical configurations \cite{R46}. The classical master equation is obtained from Eq.(3) after adiabatic elimination of coherences and reads
\begin{equation}
\frac{d P_n}{dt}=\sum_{m=1}^{N} W_{n,m} P_m(t)  
\end{equation}
where the elements of the Markov transition matrix $W$ are given by  (Sec.S1 of \cite{supp})
\begin{equation}
W_{n,m}=- \frac{2}{\gamma} (J_n^2+J_{n-1}^2)  \delta_{n,m} +\frac{2 J_n^2}{\gamma} \delta_{n,m-1} +\frac{2 J_{n-1}^2}{\gamma} \delta_{n,m+1}.
\end{equation}

\begin{figure}[t]
    \centering
    \includegraphics[width=0.45\textwidth]{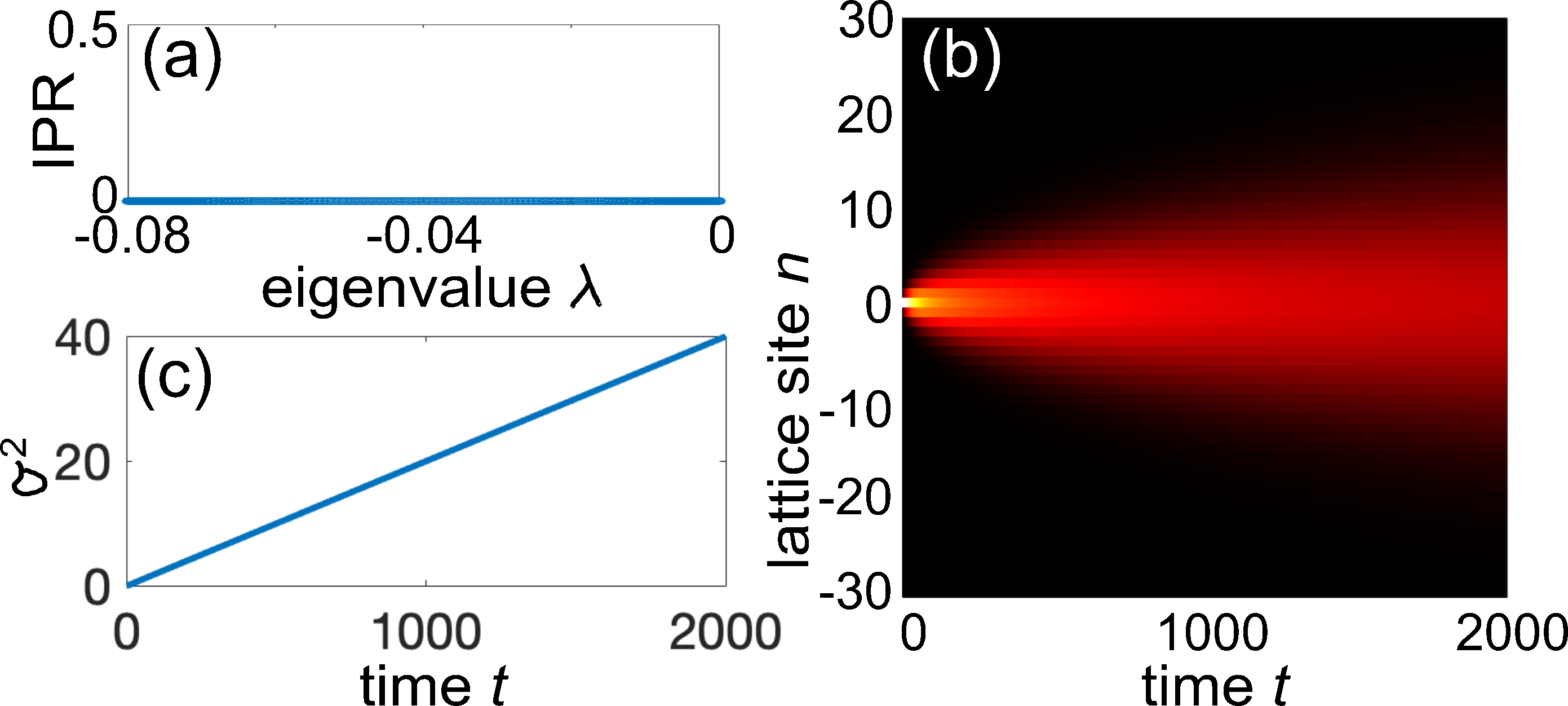}
    \caption{Suppression of ME and diffusive spreading in the generalized diagonal Aubry-Andr\'e model induced by dephasing. (a) IPR of the eigenstates of the Markov matrix $W$ versus the eigenvalues $\lambda$. Dephasing rate $\gamma=100$; other parameter values are as in Fig.1(a,b). (b,c) Spreading of an initial single-site excitation of the lattice (snapshot of $P_n(t)$ and temporal evolution of second moment $\sigma^2$). Note that the  spreading is diffusive, $\sigma^2(t)=Dt$, with diffusion coefficient $D=2 J^2/ \gamma$.}
    \end{figure}
    \begin{figure*}
    \centering
    \includegraphics[width=0.9 \textwidth]{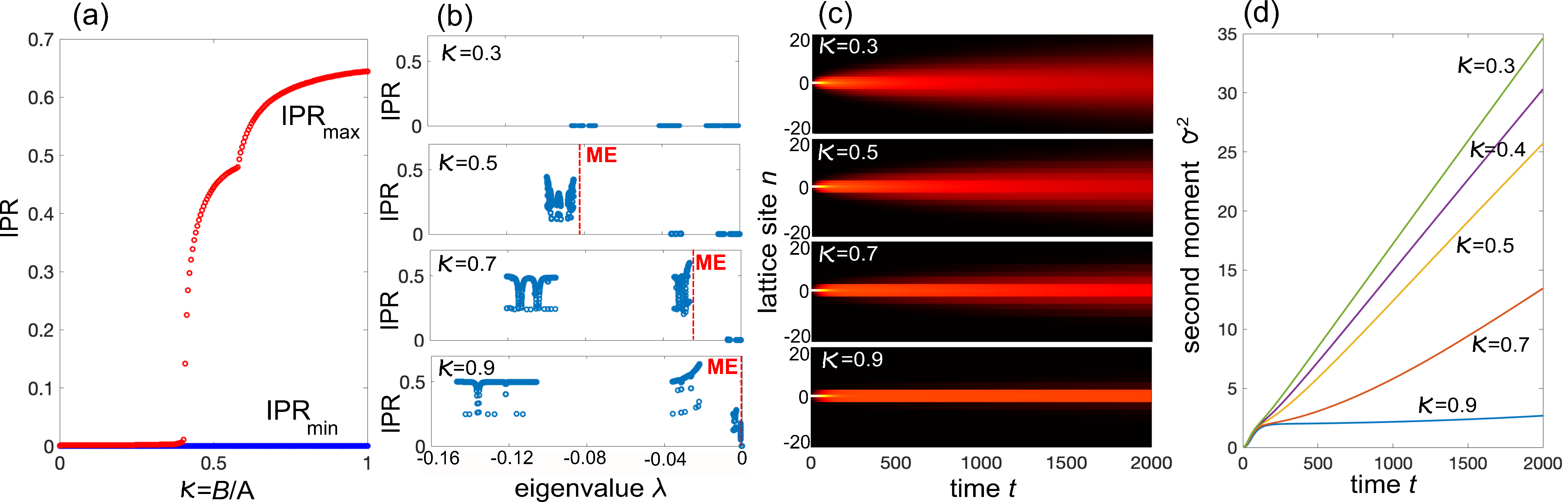}
    \caption{Dephasing-induced ME in the off-diagonal Aubry-Andr\'e model. (a) Behavior of IPR$_{\min}$ and IPR$_{max}$ versus the ratio $\kappa=B/A$. ME are created  for $\kappa>\kappa_c \simeq 0.4$. (b) Behavior of IPR of all eigenvectors of $W$ versus the eigenvalues $\lambda$ for a few increasing values of $\kappa$. Note that the ME $\lambda=\lambda_m$, separating extended ($ \lambda> \lambda_m$) from localized ( $\lambda< \lambda_m$) states approaches zero as $\kappa$ is increased toward $\kappa=1$; parameter values are $A=1$ and $\gamma=100$. (c,d) Spreading dynamics of initial single-site excitation of the lattice (snapshots of $P_n(t)$ and temporal evolution of second moment $\sigma^2$). Lattice size $L=987$.}
   \end{figure*}
The same classical master equation can be derived from the single-particle Schr\"odinger equation with periodic phase randomization of the wave function at time intervals $\Delta t= 2/ \gamma$  (Sec.S2 of \cite{supp}). Note that, since $W$ is an Hermitian matrix, the classical master equation (4) can be viewed as a Schr\"odinger equation with a Wick-rotated anti-Hermitian Hamiltonian $H'=iW$. Let us indicate by $\lambda_l$ and $\Theta_n^{(l)}$ the eigenvalues and corresponding eigenvectors of $W$ ($l=1,2,...,L$). Since $W$ is Hermitian, the eigenvalues are real and can be ordered such that $ \lambda_L \leq  ... \leq \lambda_2 \leq \lambda_1$, with the constraint $\lambda_l \leq 0$.  {Most of the eigenvectors of $W$, those with non-vanishing eigenvalues, are not physical states since they do not conserve the total probability \cite{Larson}, yet they provide a suitable complete basis for the dynamics.} There is always a vanishing eigenvalue, $\lambda_1=0$, with eigenvector $\Theta_n^{(1)}=(1/L)(...1,1,1,...)^T$. This physical state does not decay in time and  corresponds to an extended  (maximally-mixed) state. Since any initial state of the system has a non-vanishing projection onto such an eigenstate, in the long-time limit any initial localized excitation tends to spread and homogenize in all sites of the lattice with a diffusive dynamics \cite{Lars2}. However, the relaxation dynamics can be significantly slowed down when there are localized eigenvectors of $W$ with extremely long lifetimes. 
An interesting property of the matrix $W$ is that its elements do not depend on the on-site potential $V_n$. Therefore, when the disorder is introduced in the on-site potential $V_n$ solely and the hopping rates $J_n=J$ are homogeneous, such as in model (i), the incoherent hopping terms are homogeneous, all eigenvectors of $W$ are extended and thus there are not mobility edges. This means that, as expected, dephasing destroys ME and leads to delocalization. The spreading in the lattice is diffusive, i.e. $\sigma^2(t)=Dt$, with a diffusion coefficient $D=2 J^2/ \gamma$. Washing out of ME in model (i) due dephasing and diffusive spreading of excitation in the lattice is illustrated in Fig.2.\\
A very different behavior can be found in model (ii), where incommensurate disorder is introduced in the hopping rates $J_n$. In this case, $W$ displays {\em both} diagonal and off-diagonal  disorder, and numerical computation of IPR of the eigenstates of $W$ clearly shows that ME are created when the ratio $\kappa=B/A$ becomes larger than the critical value $\kappa_c \simeq 0.4$; see Fig.3(a). The figure depicts the values of  the smallest (${\rm IPR}_{min}$) and largest  (${\rm IPR}_{max}$) values of the IPR of all eigenvectors of $W$, as $\kappa$ is increased from 0 to 1. The $\kappa>1$ regime is not considered since in this case $J_n$ can vanish at some values of $n$ and there is trivial localization in the system. 
{For $ \kappa>\kappa_c$, coexistence of exponentially-localized and extended (ergodic) states, suggested by the behavior of IPR shown in Fig.3(a), is demonstrated by a finite-size scaling analysis and Lyapunov exponent computation, which are presented in Sec.S3 of \cite{supp}. The appearance of narrow region of critical states near the ME is also suggested by level statistics distribution, which shows typical level clustering behavior \cite{rs12,rs13,rs14}}.
Remarkably, when $\kappa$ approaches 1 from below, the position of the ME $\lambda_m$ moves toward zero, indicating that a large portion of localized eigenstates displays an extremely long lifetime, as shown in Figs.3(b).  The appearance of such strongly-localized eigenstates of $W$ with long lifetimes have a great impact on the dynamical spreading of excitations in the lattice, since they can trap the excitation for long times and can be thus largely slow down thermalization toward the maximally-mixed state. The transient trapping of excitation as $\kappa$ is increased toward 1 is clearly illustrated in Figs.3(c) and (d). {By diagonalization of the Liouvillian superoperator $\mathcal{L}$, we checked that the appearance of ME in the off-diagonal Aubry-Andr\'e model  persists for finite dephasing rates $\gamma$ (Sec.S4 of \cite{supp}).}\\

\begin{figure*}
    \centering
    \includegraphics[width=0.9 \textwidth]{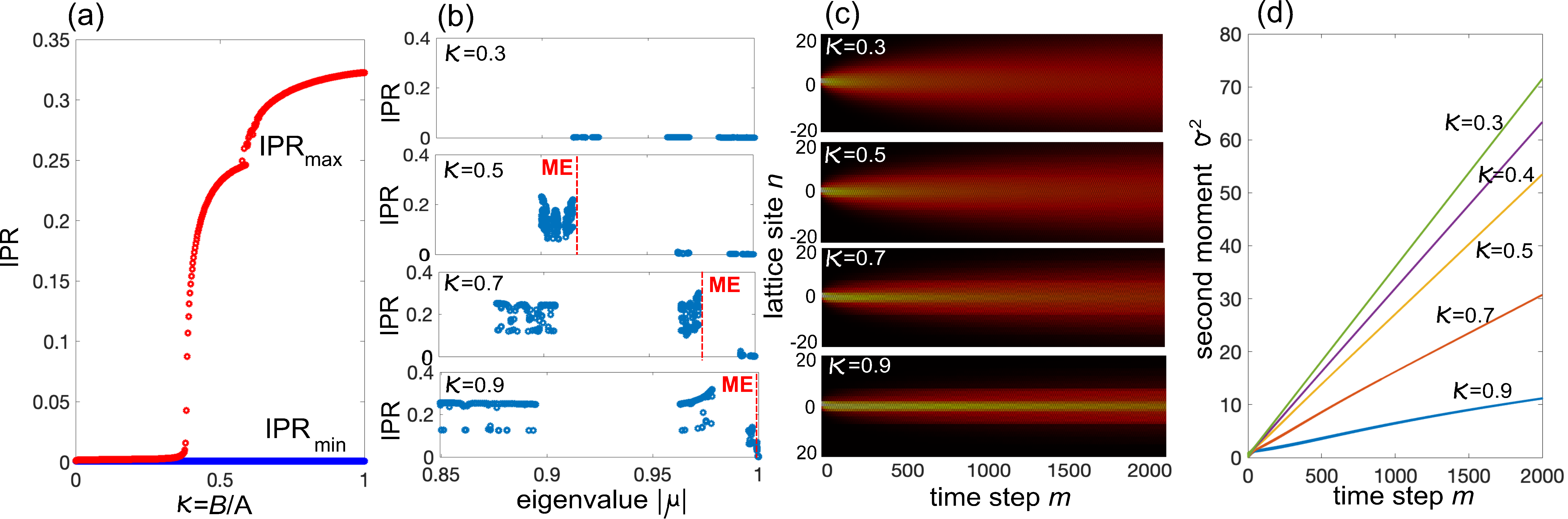}
    \caption{Dephasing-induced ME in photonic quantum walks. (a) Behavior of IPR$_{\min}$ and IPR$_{\rm max}$ versus the ratio $\kappa=B/A$ for $A=0.1$ and $\alpha=( \sqrt{5}-1)/2$. The coupling angle is $\beta_n= \pi/2-2A-2B \cos (2 \pi \alpha n)$. (b) Behavior of IPR of all eigenvectors of the incoherent matrix $\mathcal{U}$ versus $| \mu|$ for a few increasing values of $\kappa$. (c,d) Spreading dynamics of initial single-site excitation of the lattice. Panels in (c) show snapshots of $P_n^{(m)}=X_{n}^{(m)}+Y_{n}^{(m)}$ versus discrete time step $m$ for initial excitation $X_n^{(0)}=\delta_{n,0}$, $Y_n^{(0)}=0$. Panel (d) depicts the corresponding evolution of second moment $\sigma^2$. Lattice size $L=377$.}
    \end{figure*}

{\it Dephasing-induced mobility edges in photonic quantum walks.} Discrete-time quantum walks provide experimentally-accessible models to test disorder and decoherence phenomena \cite{R61,R62,R63,R64,R65,R66,R67,R67b,R68,R69}. In particular, photonic quantum walks realize synthetic lattices with controllable disorder and decoherence \cite{R65,R66,R67,R67b}, which could provide a platform for the observation of dephasing-induced ME.  We consider a photonic implementation of a quasicrystal based on light pulse dynamics in synthetic mesh lattices \cite{R70,R71,R72,R73,R74,R75}, where dephasing is emulated by random dynamic phase changes \cite{R65,R66,R67}. The system  consists of two fiber loops of slightly different lengths that are connected by a fiber coupler with a time-dependent coupling angle $\beta$. A phase modulator is placed in one of the two loops, which  impresses stochastic phases to the traveling pulses emulating dephasing effects. Off-diagonal disorder is introduced  by proper temporal modulation of the coupling angle $\beta$ between
the two fiber loops \cite{R73}.
Light dynamics is described by the set of discrete-time equations \cite{R73}
 \begin{eqnarray}
 u^{(m+1)}_n & = & \left(   \cos \beta_{n+1} u^{(m)}_{n+1}+i \sin \beta_{n+1} v^{(m)}_{n+1}  \right)  \exp (i\phi_{n}^{(m)}) \;\; \; \;\; \;   \\
 v^{(m+1)}_n & = &    i \sin \beta_{n-1} u^{(m)}_{n-1}+\cos \beta_{n-1} v^{(m)}_{n-1}  
 \end{eqnarray}
 where $u_n^{(m)}$ and $v_n^{(m)}$ are the pulse amplitudes at discrete time step $m$ and lattice site $n$ in the two fiber loops, $\beta_n$ is the site-dependent coupling angle, and $\phi_n^{(m)}$ are uncorrelated stochastic phases with uniform distribution in the range $(-\pi, \pi)$. For coupling angles $\beta_n$ close to $\pi/2$ and under coherent dynamics, i.e. for $\phi_n^{(m)}=0$, the model described by Eqs.(6) and (7) reproduces the off-diagonal Aubry-Andr\'e model with hopping rates $J_n= \pm (1/2) \cos \beta_{n+1}$ (Sec.S5 of \cite{supp}),; the model (ii) is thus retrieved after letting $\beta_n= \pi/2-2A-2B \cos ( 2 \pi \alpha n)$ ($A,B \ll 1$). When the stochastic phases are applied at every time step, the incoherent dynamics is described by the probability-conserving map (Sec.S5 of \cite{supp})
 \begin{eqnarray}
X_n^{(m+1)} & = &  \cos^2 \beta_{n+1} X_{n+1}^{(m)}+ \sin^2 \beta_{n+1} Y_{n}^{(m)} \\
Y_n^{(m+1)} & = & \sin^2 \beta_{n} X_{n}^{(m)}+ \cos^2 \beta_{n} Y_{n-1}^{(m)} 
\end{eqnarray}
  for the light pulse intensities $X_n^{(m)}=\overline{|u_n^{(m)}|^2}$ and $Y_n^{(m)}=\overline{|v_{n+1}^{(m)}|^2}$, where the overline denotes statistical average. The incoherent pulse dynamics is thus fully captured by the eigenvalues $\mu$ and corresponding eigenstates of the incoherent propagation matrix $\mathcal{U}$, defined by Eqs.(8) and (9). The eigenvalues $\mu$ satisfy the constraint $| \mu |\leq 1$, and there is always the eigenvalue $\mu_1=1$ corresponding to the non-decaying extended eigenstate $X_n=Y_n=1/(2L)$, where the photon is found with equal probability in each site of the two loops. This means that asymptotically any initially localized excitation in the lattice spreads to reach a uniform distribution. However, if there are localized eigenstates of the matrix $\mathcal{U}$ with extremely long lifetimes, i.e. with $|\mu|$ very close to 1, the spreading can be greatly slowed down and excitation can be transiently trapped in the lattice. The appearance of ME induced by dephasing effects and corresponding slowing-down of light spreading in the quantum walk is illustrated in Fig.4. This behavior is clearly similar to the one shown in Fig.3, given that for $\beta_n \sim \pi/2$ the incoherent photon dynamics can be mapped into the master equation (4) (Sec.S6 of \cite{supp}). A similar behavior is nevertheless observed even when the angles $\beta_n$ are not close to $\pi/2$.   \\ 
  
  {\it Conclusions.} In summary, we predicted that in certain one-dimensional lattices with aperiodic order ME can be created (rather than destroyed) by dephasing effects. While Anderson localization and ME arising from diagonal (on-site potential) disorder is always spoiled out by dephasing, in models where disorder is off-diagonal (i.e. in the coupling constants) ME, separating localized from extended states, can be created by dephasing effects, even when the coherent Hamiltonian does not display any Anderson localization and all states are extended. While the fate of dephasing is always to drive the system toward an incoherent state with excitation uniformly distributed in the lattice via a diffusive process, the spreading dynamics can be greatly slowed down when the Anderson-localized  states of the Markov matrix have extremely long lifetimes. The present study unveils a distinctive physical mechanics of ME formation and indicates that, contrary to a common wisdom, dephasing effects can slow down (rather than enhance) transport in a disordered lattice, a result that could be of relevance in different physical, chemical or biological systems where disorder and dephasing noise play a crucial role.

  \textit{Acknowledgments.}
The author acknowledges the Spanish State Research Agency, through the Severo Ochoa
and Maria de Maeztu Program for Centers and Units of Excellence in R\&D (Grant No. MDM-2017-0711).

\newpage

\renewcommand{\thesection}{\Alph{section}}
\renewcommand{\thefigure}{S\arabic{figure}}
\renewcommand{\thetable}{S\Roman{table}}
\setcounter{figure}{0}
\renewcommand{\theequation}{S\arabic{equation}}
\setcounter{equation}{0}

\begin{widetext}

\section{Dephasing-induced mobility edges in quasicrystals:\\
Supplemental Material}

In this Supplemental Material, we provide some technical details and analytical derivations of results presented in the main manuscript.

\subsection{S1. Derivation of the classical Markov master equation}
In this section we derive the classical master equation and explicit form of the Markov transition matrix $W$, as given by Eqs.(4) and (5) of the main text, in the limit of a large dephasing rate.\\
In the single-particle sector of Hilbert space, the Lindblad master equation yields the following evolution equations for the density matrix elements $\rho_{n,m}$ \cite{ME1,ME2}
 \begin{equation}
 \frac{d \rho_{n,m}}{dt'}  =  (i/ \gamma) (J_{n-1} \rho_{n-1,m}+J_n \rho_{n+1,m}) 
- (i/ \gamma) (J_m \rho_{n,m+1}+J_{m-1} \rho_{n,m-1}) 
   + (i/ \gamma) (V_m-V_n) \rho_{n,m}-(1- \delta_{n,m}) \rho_{n,m}  \nonumber
 \end{equation}
 where $t'= \gamma t$ is the dimensionless time scaled to the coherence time $1/ \gamma$. 
{\color{black}  When the decay rate $\gamma$ of coherences is much larger than any terms $|J_n|$ and $|V_n|$, the quantities $|J_n| / \gamma$ and $|V_n| / \gamma$ on the right hand side of the above equation are small, and we highlight this fact by writing the equation in the form
  \begin{equation}
 \frac{d \rho_{n,m}}{dt'}  =  \epsilon (i/ \gamma) (J_{n-1} \rho_{n-1,m}+J_n \rho_{n+1,m}) 
-  \epsilon (i/ \gamma) (J_m \rho_{n,m+1}+J_{m-1} \rho_{n,m-1}) 
   + \epsilon (i/ \gamma) (V_m-V_n) \rho_{n,m}-(1- \delta_{n,m}) \rho_{n,m}  
 \end{equation}
 where $\epsilon$ is a flag that marks the order of magnitude of such small quantities. In such form,
 one can solve perturbatively the equation by a standard multiple-time scale method (see for instance \cite{multiple}), and at the end of the calculation we can set $\epsilon=1$.}
 After introduction of the  multiple time scales $T_0=t'$, $T_1= \epsilon t'$, $T_2= \epsilon^2 t'$, ... we can search for a solution of Eq.(S1) as a power series expansion in $\epsilon$, i.e.
 \begin{equation}
 \rho_{n,m}(t')=\rho_{n,m}^{(0)}(t')+ \epsilon \rho_{n,m}^{(1)}(t')+ \epsilon^2 \rho_{n,m}^{(2)}(t')+...
 \end{equation}
 {\color{black} The power series expansion is meaningful provided that each term in the expansion does not show secularly growing terms in time, which would prevent the validity of Eq.(S2) after a short evolution time. In order to avoid the appearance of secular terms, solvability conditions should be satisfied, which will provide the dynamical evolution of the leading-order density matrix elements $\rho_{n,m}^{(0)}$ on the different time scales $T_0$, $T_1$, $T_2$, ...}.
Substitution of Eq.(S2) into Eq.(S1) and using the derivative rule $d/dt'= \partial_{T_0}+ \epsilon \partial_{T_1}+\epsilon^2 \partial_{T_2}+...$, after collecting the terms of the same order in $\epsilon$ a hierarchy  of equations for successive corrections to $\rho_{n,m}(t')$ is obtained. At leading order $\sim \epsilon^0$ one has
\begin{equation}
\frac{\partial \rho_{n,m}^{(0)}}{\partial T_0}=-(1-\delta_{n,m}) \rho_{n,m}^{(0)}.
\end{equation}
{\color{black} For $n=m$, the solution to Eq.(S3), which does not display secular term growing as $\sim T_0$, is simply given by  
$\rho_{n,n}^{(0)}=P_n(T_1,T_2,...)$, i.e. it is independent of $T_0$. This means that the population dynamics evolves on the slow time scales $T_1$, $T_2$, ..., i.e. the relaxation toward the stationary maximally-mixed state of the Liouvillian occurs on such slow time scales. For $n \neq m$, the solution to Eq.(S3) is given by  $\rho_{n,m}^{(0)}=\rho_{n,m}^{(0)}(0) \exp(-T_0)$, indicating that any initial coherence in the system is rapidly damped on the fast time scale $T_0$. Since the relaxation dynamics of populations occurs on the slow time scales, we can thus disregard such a fast transient decay of coherences, and let at leading order}
\begin{equation}
\rho_{n,n}^{(0)}=P_n(T_1,T_2,...) \; ,\;\; \rho_{n,m}^{(0)}=0 \; \; (n \neq m). 
\end{equation}
{\color{black} The change of populations $P_n$ occurs on the longer time scales $T_1$, $T_2$, ... due to incoherent hopping mediated by the small values of coherences at higher orders, and can be calculated from the solvability conditions by pushing the analysis up to order  $\sim \epsilon^2$.}\\
At order $\sim \epsilon$ one obtains
\begin{equation}
\partial_{T_0} \rho_{n,m}^{(1)}+(1-\delta_{n,m}) \rho_{n,m}^{(1)}=G_{n,m}^{(1)}
\end{equation} 
where we have set
\begin{equation}
G_{n,m}^{(1)} \equiv (i/ \gamma) (J_{n-1} \rho_{n-1,m}^{(0)}+J_n \rho_{n+1,m}^{(0)})
- (i/ \gamma) (J_m \rho_{n,m+1}^{(0)}+J_{m-1} \rho_{n,m-1}^{(0)}) 
   + (i/ \gamma) (V_m-V_n) \rho_{n,m}^{(0)} - \frac{\partial \rho_{n,m}^{(0)}}{\partial T_1}
\end{equation}
The solution to Eq.(S5) reads
\begin{eqnarray}
\rho_{n,n+1}^{(1)} & = & \frac{iJ_n}{\gamma}(P_{n+1}-P_{n}) \nonumber  \\
\rho_{n,n-1}^{(1)} & = & -\frac{iJ_{n-1}}{\gamma}(P_{n}-P_{n-1}) \\
\rho_{n,m}^{(1)} & = & 0 \; \; m \ne n \pm 1 \nonumber
\end{eqnarray}
with the solvability conditon
\begin{equation}
\frac{ \partial P_n}{\partial T_1}=0.
\end{equation}
Equation (S7)  provides the leading-order terms ($ \sim \epsilon$) of non-vanishing coherences in the system, which occur along the two diagonals $m= n \pm 1$.\\
At order $\sim \epsilon^2$ one finally obtains
\begin{equation}
\partial_{T_0} \rho_{n,m}^{(2)}+(1-\delta_{n,m}) \rho_{n,m}^{(2)}=G_{n,m}^{(2)}
\end{equation} 
where we have set
\begin{equation}
G_{n,m}^{(2)} \equiv (i/ \gamma) (J_{n-1} \rho_{n-1,m}^{(1)}+J_n \rho_{n+1,m}^{(1)})
- (i/ \gamma) (J_m \rho_{n,m+1}^{(1)}+J_{m-1} \rho_{n,m-1}^{(1)}) 
   + (i/ \gamma) (V_m-V_n) \rho_{n,m}^{(1)} - \frac{\partial \rho_{n,m}^{(0)}}{\partial T_2}
\end{equation}
In particular, if we specialize Eq.(S9) for the populations ($m=n$), one obtains the solvability condition $G_{n,n}^{(2)}=0$, i.e.
\begin{equation}
\frac{\partial P_n}{\partial T_2} = (i/ \gamma) (J_{n-1} \rho_{n-1,n}^{(1)}+J_n \rho_{n+1,n}^{(1)})
- (i/ \gamma) (J_n \rho_{n,n+1}^{(1)}+J_{n-1} \rho_{n,n-1}^{(1)}).
\end{equation}
Substitution of Eq.(S7) into Eq.(S11) yields
\begin{equation}
\frac{\partial P_n}{\partial T_2}= -\frac{2}{\gamma^2}(J_n^2+J_{n-1}^2)P_n+\frac{2 J_n^2}{\gamma^2} P_{n+1}+\frac{2 J_{n-1}^2}{\gamma^2} P_{n-1}.
\end{equation}
Once the solvability condition Eq.(S12) is satisfied, the higher-order corrections to coherences, at order $\sim \epsilon^2$, can be obtained by solving Eq.(S9).\\
If we stop the analysis at this order $\sim \epsilon^2$, using the derivative rule $dP_n/dt= \gamma (\partial_{T_0} P_n+\epsilon \partial_{T_1}P_n+ \epsilon^2 \partial_{T_2}P_n)= \gamma \epsilon^2 \partial_{T_2}P_n$ in physical time $t$, after letting $\epsilon=1$ from Eq.(S12) one then obtains Eqs.(4) and (5) given in the main text.


\subsection{S2. Derivation of the classical master equation from the Schr\"odinger equation with periodic phase randomization}
Let us consider the single-particle coherent evolution in the tight-binding lattice and let us assume that at every time interval $\Delta t$ the phase of the wave function $\psi_n$ at any lattice site $n$ is randomized. Such a randomized phase process basically emulates dephasing effects in the dynamics, and provides a simple tool to experimentally emulate dephasing phenomena in photonic quantum walks \cite{S1}. 
 The time evolution of the wave function amplitudes $\psi_n(t)$ read
 \begin{equation}
 i \frac{d \psi_n}{dt}= \sum_m H_{n,m} \psi_m+ \psi_n \sum_{\alpha=1,2,3,...} \varphi_n^{(\alpha)} \delta(t-\alpha \Delta t)
 \end{equation}
where $H_{n,m}$ are the matrix elements of the single-particle tight-binding Hamiltonian $H$ and the last term on the right hand sides of Eq.(S13) describes the dephasing process. Here $\varphi_n^{(\alpha)}$ are assumed to be uncorrelated stochastic phases, both in site index $n$ and time step $\alpha$, with a given probability density function. Fully coherent dynamics is obtained by letting $\varphi_n^{(\alpha)}=0$, whereas fully incoherent (classical) dynamics is obtained by assuming a uniform distribution in the range $(-\pi, \pi)$ for the probability density function. 
Under fully coherent dynamics, the wave function amplitudes evolve according to $ \psi_n(t)=\sum_m U_{n,m}(t) \psi_m(0)$, where $U(t)=\exp(-i Ht)$ is the coherent propagator. On the other hand, for fully incoherent dynamics indicating by $P_n(t_\alpha)={\overline {|\psi_n(t_{\alpha})|^2}}$ the occupation probabilities (populations) at various sites of the lattices, where $t_{\alpha}=\alpha \Delta t$ and the overbar denotes statistical average, the time evolution is described by the classical map
\begin{equation}
P_n(t_{\alpha+1})=\sum_m \mathcal{U}_{n,m}P_m(t_{\alpha})
\end{equation}
where $\mathcal{U}_{n,m}=|U_{n,m} (\Delta t)|^2$ in the incoherent propagator. Equation (S14) can be readily obtained by solving Eq.(S13) in each time interval $\Delta t$ and then taking the statistical average. The classical master equation (4) given in the main text is finally obtained in the small $\Delta t$ limit. In fact,  assuming a short time interval $\Delta t$ between successive stochastic phases, one can expand the coherent propagator in power series of $\Delta t$, yielding $U(\Delta t) \simeq 1-i \Delta t H-(1/2) \Delta t^2 H^2$ and thus
\begin{equation}
\mathcal{U}_{n,m}= \delta_{n,m}+ \Delta t^2 \left(  |H_{n,m}|^2 -\delta_{n,m} \sum_l |H_{n,l}|^2 \right)
\end{equation}
up to the order $ \sim \Delta t^2$.  Since $P_n(t_{\alpha})$ evolves slowly with index $\alpha$, i.e. after each short time interval $\Delta t$, one can set $P_n(t+\Delta t)=P_n(t)+(dP_n/dt)$ and consider $t_{\alpha}=t$ a continuous variable. From Eqs.(S14) and (S15) one then obtains the classical master equation
\begin{equation}
\frac{d P_n}{dt}=\sum_{m=1}^{N} W_{n,m} P_m(t)  
\end{equation}
where the elements of the Markov transition matrix $W$ are given by
\begin{equation}
W_{n,m}= \Delta t \left(  |H_{n,m}|^2 -\delta_{n,m} \sum_l |H_{n,l}|^2 \right)= \left\{
\begin{array}{cc}
-\Delta t \sum_{l \neq n} |H_{n,l}|^2 & n=m \\
\Delta t |H_{n,m}|^2 & n \neq m
\end{array}
\right.
\end{equation}
For the tight-binding model describing the quasicrystal, one has $H_{n,m}=V_n \delta_{n,m}+J_{n} \delta_{n,m-1}+J_{n-1} \delta_{n,m+1}$, so that one obtains
\begin{equation}
W_{n,m}= \left\{
\begin{array}{cc}
-\Delta t (J_{n}^2+J_{n-1}^2)  & n=m \\
\Delta t J_{n}^2 & n=m-1 \\
\Delta t J_{n-1} & n=m+1 \\
0 & n \neq m, m \pm 1
\end{array}
\right.
\end{equation}
which reduces to Eq.(5) of the main text provided that we assume $\Delta t = 2 / \gamma$. 

{\color{black}{
\subsection{S3. Finite-size scaling analysis of localization, Lyapunov exponent and level spacing statistics}

The localization properties of a wave function $\psi_n^{(l)}$ with energy $E_{l}$, normalized as $ \sum_{n=1}^{L} | \psi_n^{(l)}|^2=1$, are characterized by the generalized inverse
participation ratio, defined by
\begin{equation}
{\rm  IPR}_{l}^{(q)}  \sum_{n=1}^{L} | \psi_n^{(l)} |^{2q},
 \end{equation}
 where $q>0$ is a positive number,   and by the exponent $\beta_{l}^{(q)}$, which is given by (see for instance \cite{r1,r2,r3,r4,r5})
  \begin{equation}
  \beta_{l}^{(q)}= \lim_{L \rightarrow \infty} \frac{\ln {\rm IPR}_{l}^{(q)}}{\ln (1/L)} .
    \end{equation}
   The quantity $D_l(q)= \beta_i^{(q)}/(q-1)$ is known as the {\it fractal dimension}. Note that  for $q=2$ one has $D_l(q)= \beta_l^{(q)}$. For extended (ergodic) and localized states one has $
   D_l(q)=1$ and $D_l(q)=0$, respectively, whereas any other behavior of $D_l(q)$ implies multifractality \cite{r5}, i.e. critical states. In the finite-size scale analysis of localization,  the golden ratio $\alpha=(\sqrt{5}-1)/2$ is  approached by the Fibonacci numbers via the relation $\alpha= \lim_{l \rightarrow \infty} F_{l-1}/F_l$, where the Fibonacci numbers $F_l=1, 1, 2, 3, 5, 8, 13, 21, 34, 55, 89, 144, 233, 377,$ $610, 987, 1597, 2584, 4181, 6765, 10946, 17711, ... $ are defined recursively
by $F_{l+1}= F_{l-1}+F_l$, with $F_0= F_1=1$. Numerically, we successively change the system size $L=F_l$ to approach the irrational number, and extrapolate IPR$_{l}^{(q)}$ and $\beta_{l}^{(q)}$ by their asymptotic behavior as $L$ is increased. Open boundary conditions are assumed in the simulations, with system size $L$ typically spanning the range of Fibonacci numbers from $L=34$ to $L=17711$.\par
\begin{figure}
\includegraphics[width=0.5\textwidth]{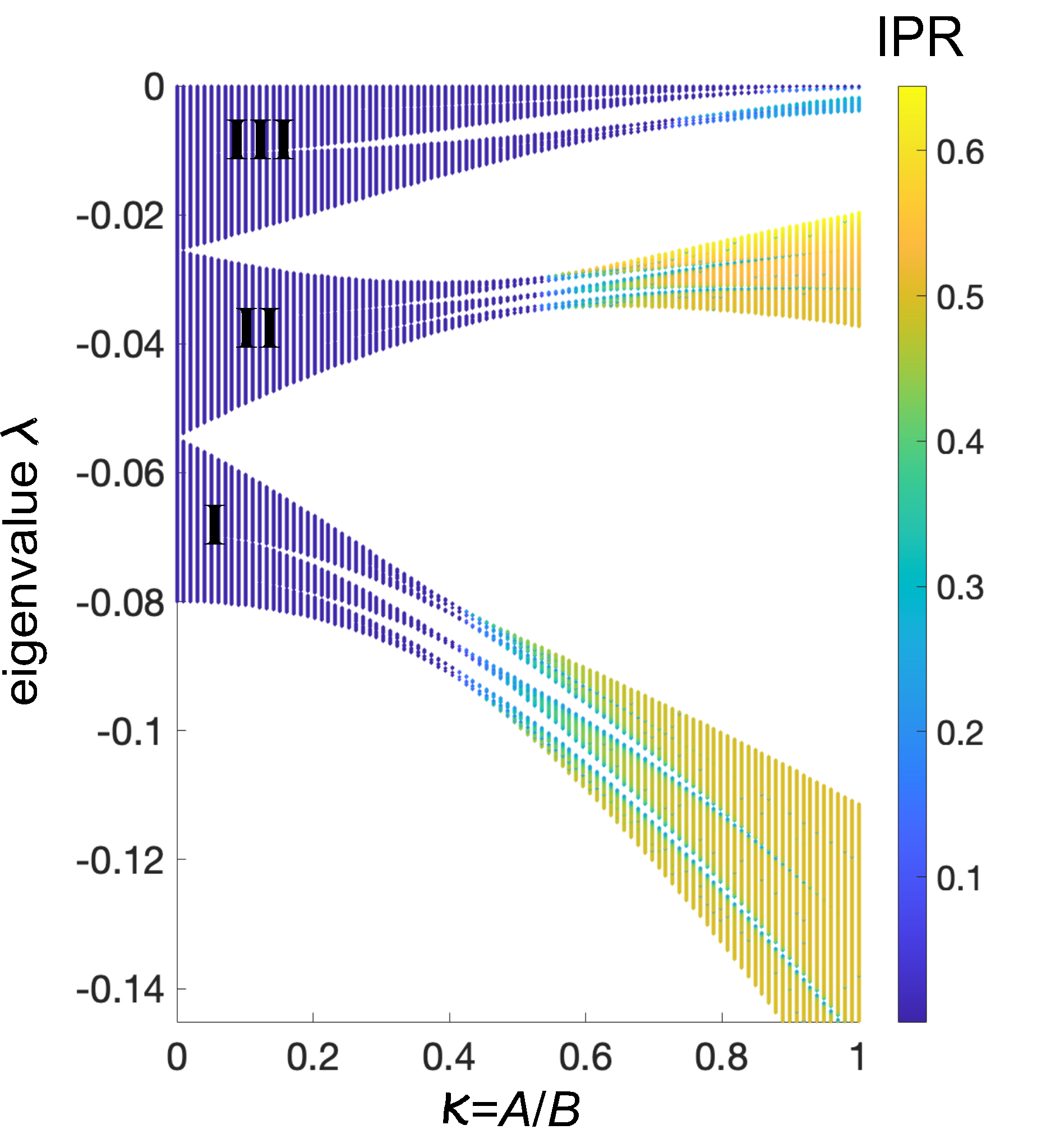}
\caption{ \color{black}{{Energy spectrum $\lambda$ versus $\kappa=B/A$ for the Markov transition matrix $W$ in the off-diagonal Aubry-Andr\'e model with dephasing. The energy spectrum (in units of $A$) has been numerically computed by diagonalization of $W$ on a lattice of size $L=4181$ with open boundary conditions. The different colors in the plot relate to different values of IPR of corresponding eigenvectors. $q=2$ has been assumed in the computation of IPR. Note that the energy spectrum comprises three main pseudo bands, labelled by roman numbers I,II,III, separated by two wide gaps. As $\kappa$ is increased above the critical value $\kappa_c \simeq 0.4$, coexistence of localized and extended states, corresponding to large and small IPR values, is suggested.}}}
\end{figure}

\begin{figure}
\includegraphics[width=0.96\textwidth]{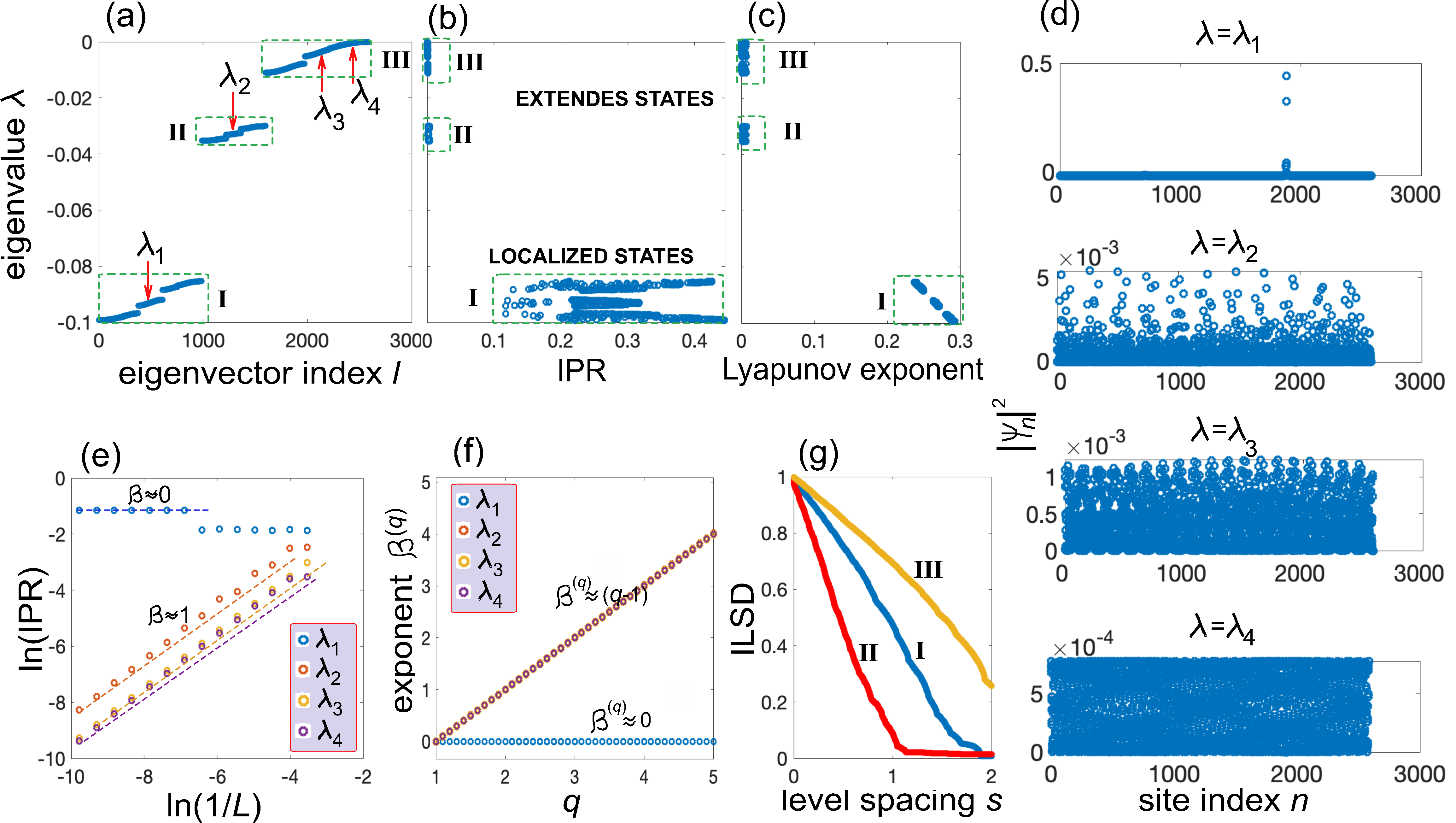}
\caption{ \color{black}{{(a) Eigenvalues $\lambda$ (energy spectrum) of the Markov transition matrix $W$ in the off-diagonal Aubry-Andr\'e model with dephasing for $A=1$ and $\kappa=B/A=0.5$. (b) IPR of the eigenvectors (lattice size $L=2584$)  with $q=2$ and (c) corresponding Lyapunov exponent (inverse of the localization length). (d) Profiles of the eigenvectors (plots of $| \psi_n^{(l)}|^2$) corresponding to the four eigenvalues $\lambda_1=-0.093$, $\lambda_2=-0.0329$, $\lambda_3=-4 \times 10^{-3}$ and $\lambda_4=-2 \times 10^{-4}$, indicated by the vertical arrows in (a). (e) Behavior of ${\rm ln}$(IPR) (with $q=2$) versus lattice size ${\rm ln} (1/L)$ for the four eigenvectors of panels (d). The estimated slopes $\beta$ of the four curves, $\beta \simeq 0$ for $\lambda=\lambda_1$ and $\beta \simeq 1$ for $\lambda=\lambda_2, \lambda_3$ and $\lambda_4$, indicate that the eigenvectors are either localized ($\lambda=\lambda_1$) or extended ergodic ($\lambda=\lambda_2, \lambda_3, \lambda_4$). (f) Behavior of exponent $\beta^{(q)}$ versus $q$ for the four eigenvectors of panels (d)  (lattice size $L=6765$). (g) Numerically-computed integrated level spacing distribution (ILSD) for the three regions I, II and III of energy spectrum. A lower cutoff $s_0=\times 10^{-6}$ has been assumed in the computation of ILSD.}}}
\end{figure}
  In the off-diagonal Aubry-Andr\'e model with dephasing, the Markov transition matrix $W$ displays incommensurate disorder in both hopping rates and on-site energies, and its energy spectrum shows a typical
Cantor-like set structure of incommensurate models  \cite{r6},  with infinitely many small bands separated by small gaps. The behavior of energy spectrum of $W$ versus $\kappa=B/A$, computed by diagonalization of the Markov matrix $W$ on a lattice of size $L=4181$, is shown in Fig.S1. The spectrum basically comprises three main pseudo bands, labelled as I,II and III in the figure, separated by large gaps. For $\kappa$ below the critical value $\kappa_c \simeq 0.4$, all eigenvectors of $W$ are extended, whereas for $\kappa>\kappa_c$ coexistence of extended and localized eigenstates in suggested from the IPR behavior, which is indicated by the different colors in the plot.\\
To get deeper insights into the localization properties of the eigenvectors and the localization-delocalization transition, extended numerical simulations have been performed computing the behavior of Lyapunov exponent (inverse of localization length), fractal dimension and integrated level spacing distribution (ILSD). Typical numerical results are shown in Figs.S2 and S3 for $\kappa$ well above ($\kappa=0.5$, Fig.S2) and just above ($\kappa=0.407$, Fig.S3) the critical point $\kappa_c \simeq 0.4$. In the former case (Fig.S2) all eigenvectors in pseudo bands I and II are localized, while in pseudo band III they are all extended, as clearly shown in Figs.S2(b) and (c). The localization of the eigenvectors is characterized by the Lyapunov exponent (inverse of the localization length), which is depicted in Fig.S2(c). The Lyapunov exponent $\gamma(\lambda_{l})$ for the eigenvector $\psi_n^{(l)}$ of $W$ with eigenvalue (energy) $\lambda_l$ has been numerically computed using the relation \cite{r7,r8,r9}
\begin{equation}
\gamma(\lambda_l)=\lim_{L \rightarrow \infty} \frac{1}{L-1} \sum_{ \lambda_n \neq \lambda_l}  \log |\lambda_n-\lambda_l|-\lim_{L \rightarrow \infty} \frac{1}{L-1} \sum_{n=1}^{L-1}  \log |W_{n,n+1}|
\end{equation}
assuming a large lattice size (typically $L=6765$). For an energy $\lambda$ in the spectrum, $\gamma(\lambda)$ vanishes for an extended or critical state, whereas it assumes a finite value for an exponentially-localized state, $ 1 / \gamma(\lambda)$ providing the characteristic localization (decay) length of the eigenvector. As shown in Fig.S2(c), all eigenvectors in spectral region I display a finite Lyapunov exponent, with a localization length that increases as $\lambda$ is increased but that remains finite over the entire spectral region. Conversely, in the pseudo bands II and III the Lyapunov exponent is very close to zero, indicating that in such regions all the eigenvectors are either extended or critical.  Since the Lyapunov exponent jumps from finite to almost vanishing values, the mobility edge for $\kappa=0.5$  falls in the wide gap separating the pseudo bands I and II. This behavior also suggests that there are not critical states, and the mobility edge separates extended (ergodic) from exponentially-localized eigenvectors. A few examples of localized and extended eigenstates, corresponding to the four eigenvalues $\lambda_1=-0.093$, $\lambda_2=-0.0329$, $\lambda_3=-4 \times 10^{-3}$ and $\lambda_4=-2 \times 10^{-4}$,  are depicted in Fig.S2(d). To check the absence of critical states, Fig.S2(e) shows the behavior of ${\rm ln}$(IPR) (with $q=2$) versus lattice size ${\rm ln} (1/L)$ for the four eigenvectors of panel (d). The estimated slopes $\beta$ of the four curves, $\beta \simeq 0$ for the eigenvector with eigenvalue $\lambda=\lambda_1$ and $\beta \simeq 1$ for the eigenvectors with eigenvalues $\lambda=\lambda_2, \lambda_3$ and $\lambda_4$, indicate that the eigenvectors are either localized ($\lambda=\lambda_1$) or extended ergodic ($\lambda=\lambda_2, \lambda_3, \lambda_4$), i.e. they do not display multifractality. The nature of the eigenstates is also confirmed by the behavior of exponents $\beta^{(q)}$ versus $q$, for a fixed value of lattice size $L$, which is depicted in Fig.S2(f). 

 A different behavior is found near the critical point (Fig.S3). In this case, besides extended and localized eigenstates, there is evidence of critical states at the mobility edge separating extended and localized states, which falls in a narrow sub-band of pseudo band I [inset of Fig.S3(a)]. The eigenvector with eigenvalue $\lambda=\lambda_3$, depicted in the third row of Fig.S3(d), is neither fully localized nor fully extended, i.e. it is a critical state displaying multifractality, as indicated by the behavior of $\beta$ for this eigenvector shown in Figs.S3(e) and (f).

\begin{figure}
\includegraphics[width=0.96\textwidth]{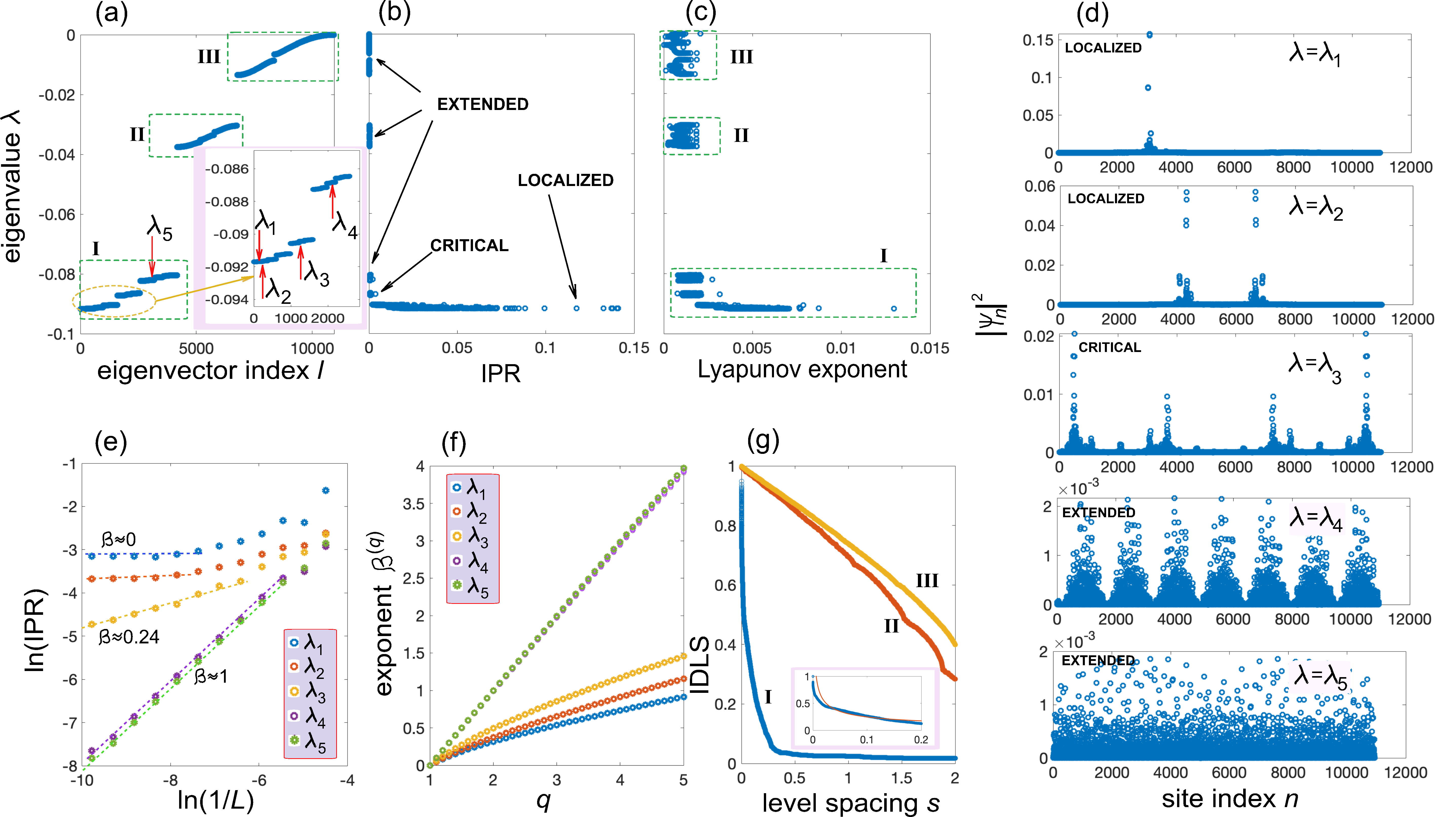}
\caption{ \color{black}{{Same as Fig.S2 but for $\kappa=0.407$. The five eigenvalues in panels (a), (d),(e) and (f) are: $\lambda_1=-0.09168$, $\lambda_2=-0.09157$, $\lambda_3=-0.09105$, $\lambda_4=-0.0868$, and $\lambda_5=-0.08107$. The inset in (g) shows an enlargement of the ILSD near $s=0$ for the spectral region I (bold blue curve), and the corresponding fitting curve $\sim s^{-1/2}$ (thin red curve).}}}
\end{figure}
   
Finally, we performed an analysis of level spacing statistics in order to check the appearance of critical states.    
   Level spacing statistics is an
important tool in the study of spectra of disordered systems, which can provide major insights into the localization properties and critical points of a system  \cite{r10,r11}.
It is well known that for a quantum system with uncorrelated disorder
localized and delocalized levels follow different
universal level spacing distributions $P(s)$ in the thermodynamic limit, defined as the probability density of level spacings $s$ of the adjacent
levels \cite{r10,r11}. In the extended phase $P(s)$ follows the Wigner surmise $P_W(s) \sim s \exp(-cs^2)$ distribution (with $c$ a constant), whereas in the localized phase $P(s)$ follows
a Poisson law $P_P(s) \sim \exp(-s)$. Thus, as disorder increases, the metal-insulating transition
is accompanied by a transition of the level spacing distributions from the Wigner surmise to
the Poisson law. The Wigner surmise is characterized by the typical level repulsion since $P(s) \rightarrow 0$ as $s \rightarrow 0$, whereas
for the Poisson distribution level repulsion is absent. In a disordered system displaying mobility edges, the level spacing distribution $P(s)$ changes discontinuously at the mobility
edge from $P_P (s)$ to $P_W(s)$. 
In models with incommensurate potentials, the most notable one being the Aubry-Andr\'e model, the energy level spacing distributions in the extended and localized phases do not follow such universal behaviors \cite{r12,r13,r14,r15,r16}. Interestingly, at the critical point, where the eigenvectors are critical, $P(s)$ is singular as $s \rightarrow 0$ and displays the inverse power law $P(s) \sim s^{-3/2}$, indicating energy level clustering (rather than repulsion). Such an inverse power law and level clustering have been suggested as signatures of critical wave functions in incommensurate potential models \cite{r13,r14,r15}. 
Owing to the fragmented structure of energy spectrum, formed by an uncountable set of levels  (like in the Cantor set), rather than computing the level spacing distributions $P(s)$ it is more convenient 
to calculate an integrated level-spacing distribution (ILSD) which, apart from normalization reads \cite{r13,r14,r15}
\begin{equation}
{\rm ILSD} (s)=\int_{s}^{\infty} ds' P(s')
\end{equation}
and provides the fraction of the total number of gaps larger than some size $s$. The ILSD can be computed in any given range of energies, either in the localized or extended regions, to characterize the localization features of the eigenvectors.
 For a given range of energies $(\mathcal{E}_1,\mathcal{E}_2)$, the level spacing $s$ is defined as $s=(E_{n+1}-E_n)/(W/N_E)$, where $E_1 \leq E_2,...,E_{N_E} $ are the $N_E$ energy levels that fall in the range $(\mathcal{E}_1, \mathcal{E}_2)$ and $W=(\mathcal{E}_2-\mathcal{E}_1)$. The ILSD can be
normalized by introducing a lower cutoff $s_0>0$ \cite{r13}. In Figs.S2(g) and S3(g) we show the numerically-computed behavior of the ILSD in the spectral intervals I,II and III defined by the three main pseudo bands. A clear different behavior is observed for the ILSD curves in region I for $\kappa=0.5$, where the ME falls in a gap and there are not critical states [Fig.S2(g)], and $\kappa=0.407$, where the ME falls in a narrow sub-band of region I [Figs.S3(g)]. In the latter case the ILSD curve displays a steep increase near $s=0$ [see inset in Fig.S3(g)], indicating level clustering. Such a result is a  signature of the emergence of critical states for $\kappa=0.407$, which are absent when $\kappa=0.5$.

\subsection{S4. Mobility edges in the off-diagonal Aubry-Andr\'e model with a finite dephasing rate}
  
The key result of dephasing-induced ME presented in the main manuscript for the off-diagonal Aubry-Andr\'e model has been demonstrated in the strong dephasing regime, which justifies a classical reduction of the quantum master equation \cite{Kamia}. Here we show that such a phenomenon persists for finite values of the dephasing rate as well, i.e. beyond the classical limit. 
This entails the diagonalization of the full Liouvillian superoperator $\mathcal{L}$ entering in the Lindblad master equation. In a lattice of size $L$ the density matrix $\rho$ in the single particle sector of Hilbert space is represented by a vector of size $L^2$ and the Liouvillian superoperator $\mathcal{L}$ by a $L^2 \times L^2$ square matrix. 
This makes the diagonalization problem computationally challenging for very large system sizes, however one can provide some insightful results of the cross-over between fully coherent and fully incoherent regimes taking a relatively large value of $L$ to emulate a quasicrystal, yet keeping the eigenvalue problem tractable with standard MatLab eig solver. In the numerical simulations, the golden mean $\alpha$ has been approximated by the Fibonacci ratio $\alpha=34/55$, corresponding to a system size $L=55$.
The eigenvalue problem reads $\mathcal{L} \rho= \lambda \rho$, and the stationary state $\rho_s$, corresponding to the vanishing eigenvalue $\lambda=0$, is the maximally-mixed states, with vanishing coherences and populations uniformly distributed over the $L$ sites of the lattice. The other eigenvectors of the Liouvillian superoperator correspond to eigenvalues $\lambda$ with negative real part, and only those with long lifetime (i.e. small values of $| {\rm Re}(\lambda)|$) are relevant for the relaxation dynamics. For a given eigenvector $\rho$ of $\mathcal{L}$, the IPR is defined for the populations as IPR$= \sum_n \rho_{n,n}^2$. An example of the spectrum of the Liouvillian superoperator $\mathcal{L}$ and corresponding behavior of IPR of all eigenvectors, for parameter values $A=1$, $B=0.7$ and $\gamma=A=1$, is given in Fig.S4. The eigenvalue spectrum of $\mathcal{L}$ comprises a majority of eigenvalues with real part grouped near $-\gamma$, corresponding to the most damped eigenvectors containing the coherences, and a set of eigenvalues that condensate near $\lambda=0$, i.e. closer to  the stationary maximally-mixed eigenvector $\rho_s$ of $\mathcal{L}$. The behavior of IPR clearly indicates that for such set of eigenvectors the populations are markedly localized in the region below the mobility edge $\lambda_m$, and extended above such a ME. Hence we have clear indications that ME appear even for weak-to-moderate values of the dephasing rate $\gamma$. To corroborate such an indication, in Fig.S5 we plot the behavior of IPR$_{\rm min}$ and IPR$_{\rm max}$ (among all eigenvectors of  $\mathcal{L}$) versus $ \kappa=B/A$ for a few increasing values of the dephasing rate $\gamma$. The figure clearly shows that there is still a phase transition at some critical value $\kappa_c$ even for weak values of the dephasing rate, with $\kappa_c$ approaching the value $\kappa_c=0.4$  in the strong dephasing (classical) regime $\gamma /A \gg 1$ and the value $\kappa_c=1$ in the weak dephasing regime $\gamma /A \ll 1$, as expected.}}

\begin{figure}
\includegraphics[width=0.8\textwidth]{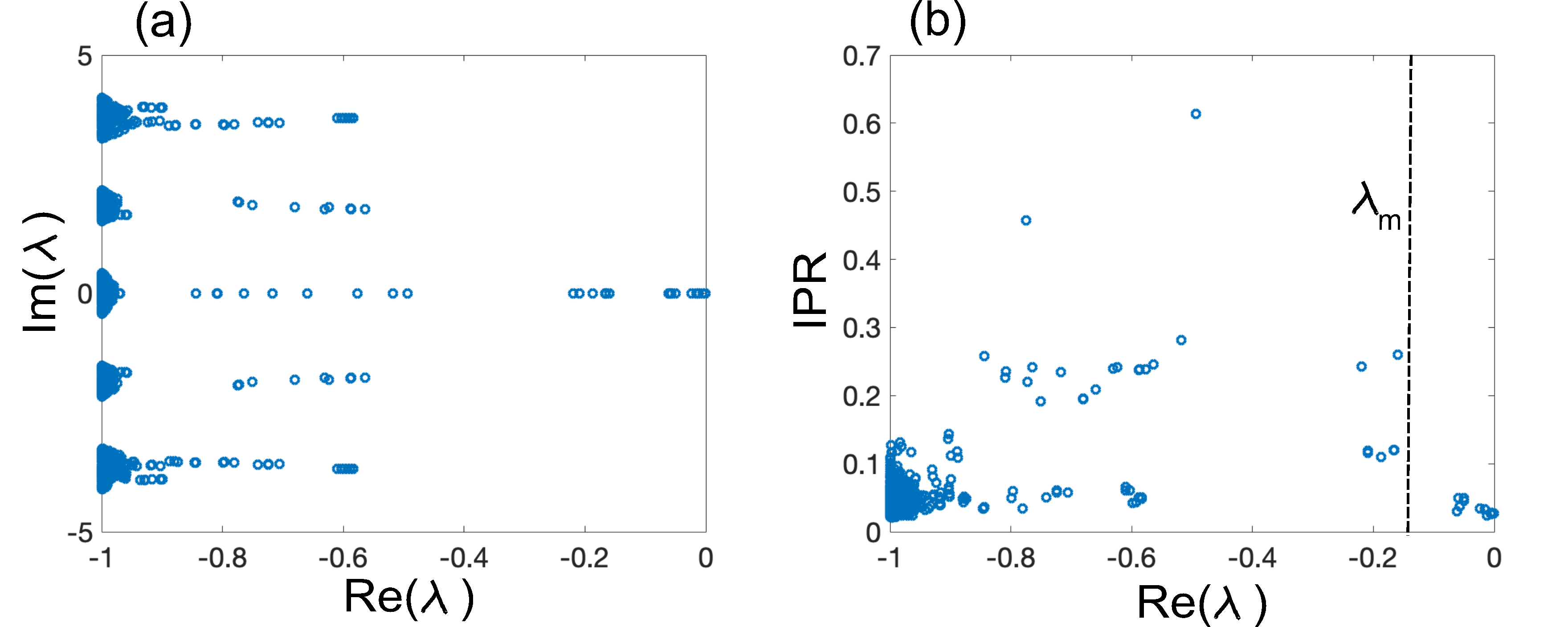}
\caption{ \color{black}{{(a) Spectrum of the Liouvillian superoperator $\mathcal{L}$ for the off-diagonal Aubry-Andr\'e model with a finite dephasing rate. Parameter values are $A=1$, $B=0.7$ and $\gamma=1$. Lattice size $ L=55$. (b) Behavior of the IPR of all eigenvectors of $\mathcal{L}$.  }}}
\end{figure}

\begin{figure}
\includegraphics[width=0.92\textwidth]{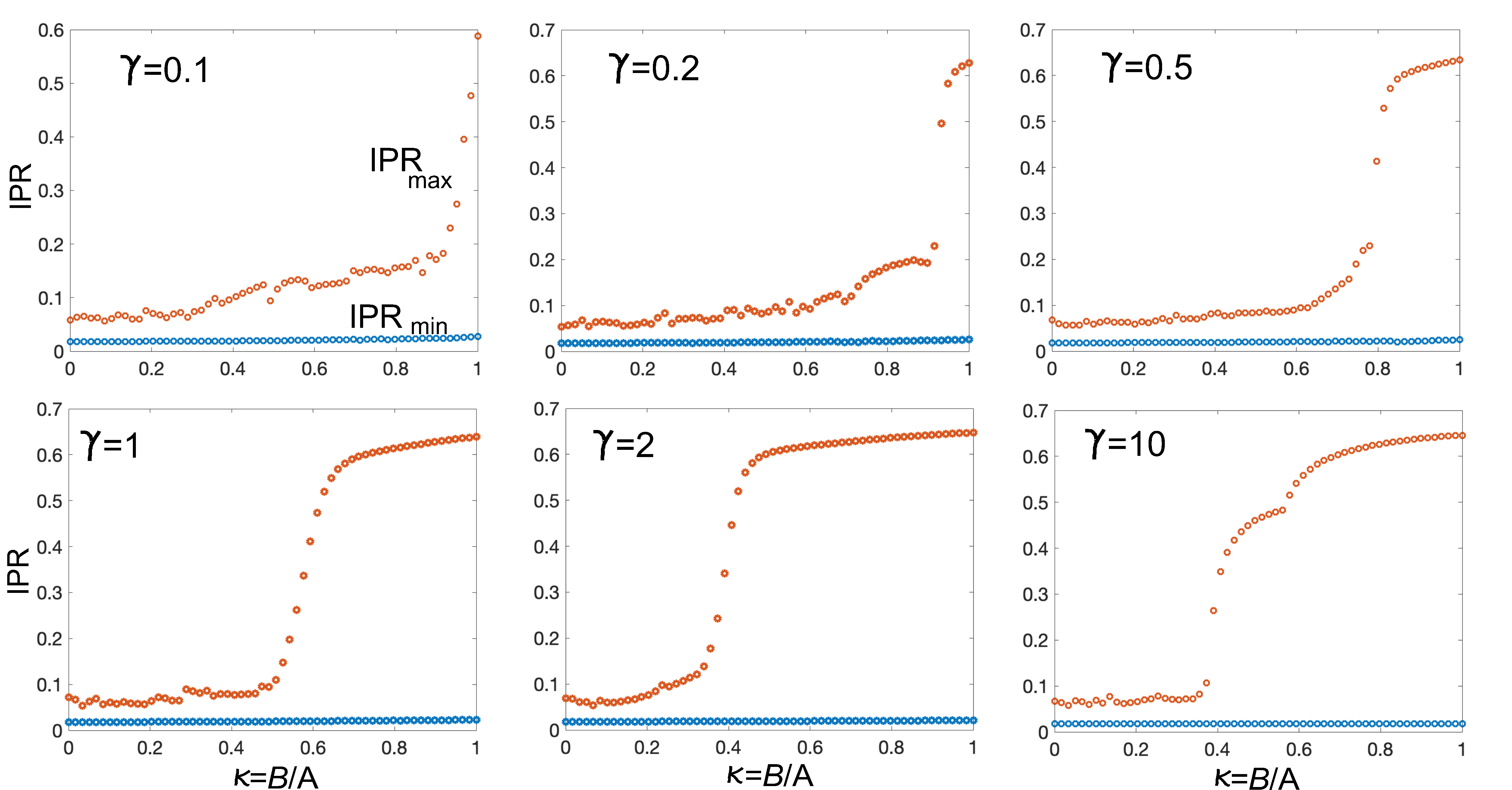}
\caption{ \color{black}{{Behavior of IPR$_{\rm min}$ and IPR$_{\rm max}$ of eigenstates of the Liouvillian superoperator $\mathcal{L}$ versus $\kappa=B/A$ for the off-diagonal Aubry-Andr\'e model at increasing values of dephasing rate $\gamma$. Parameter values are $A=1$, $L=55$.}}}
\end{figure}

\subsection{S5. Off-diagonal Aubry-Andr\'e model emulated by discrete-time photonic quantum walks}
Here we show that the off-diagonal Aubry-Andr\'e model can be emulated by the discrete-time quantum walk of photons in a synthetic mesh lattice realized by two coupled fiber loops with slightly unbalanced lengths.
The light pulse dynamics with inhomogeneous coupling angles $\beta_n$ is described by the set of discrete-time equations \cite{S2}
 \begin{eqnarray}
 u^{(m+1)}_n & = & \left(   \cos \beta_{n+1} u^{(m)}_{n+1}+i \sin \beta_{n+1} v^{(m)}_{n+1}  \right)  \exp (i\phi_{n}^{(m)}) \;\; \; \;\; \;   \label{S19}\\
 v^{(m+1)}_n & = &    i \sin \beta_{n-1} u^{(m)}_{n-1}+\cos \beta_{n-1} v^{(m)}_{n-1} \label{S20} 
 \end{eqnarray}
 where $u_n^{(m)}$ and $v_n^{(m)}$  are the pulse amplitudes at discrete time step $m$ and lattice site $n$ in the two fiber loops, $\beta_n$ is the site-dependent coupling angle, and $\phi_n^{(m)}$ are uncorrelated stochastic phases with uniform distribution in the range $(-\pi, \pi)$.

The coherent quantum walk is obtained by Eqs.(\ref{S19}) and (\ref{S20}) by letting $\phi_n^{(m)}=0$. Let us assume a coupling angle $\beta_n$ close to $\pi/2$, i.e. let us assume
\begin{equation}
\beta_n= \frac{\pi}{2}-\theta_n \label{S21}
\end{equation}
with $| \theta _n| \sim \epsilon \ll 1$. Up to order $\sim \epsilon$, we may set $\sin \beta_n \simeq 1$ and $\cos \beta_n \simeq \theta_n$ in Eqs.(\ref{S19}) and (\ref{S20}). After letting $u_n^{(m)}=(i)^m F_n^{(m)}$, elimination of the amplitudes $v_n^{(m)}$ 
from Eqs.(\ref{S19}) and (\ref{S20}) yields the following second-order difference equation for $F_n^{(m)}$
\begin{equation}
F^{(m+1)}_n=F^{(m-1)}_n-i \left( \theta_n F_{n-1}^{(m)}+\theta_{n+1}F_{n+1}^{(m)}  \right) \label{S22}
\end{equation}
which is accurate up to order $\sim \epsilon$. Equation (\ref{S22}) indicates that the amplitude $F_{n}^{(m)}$ changes by a small quantity, of order $\epsilon$, every two discrete time steps, i.e. when $m$ is increased to $(m+2)$. Therefore, the most general solution to Eq.(\ref{S22}) can be written as the superposition
\begin{equation}
F_n^{(m)}= \psi_n^{(+)}(m)+(-1)^m \psi_n^{(-)}(m) \label{S23}
\end{equation}
where the amplitudes $\psi_n^{(\pm)}(m)$ are slowly-varying functions of $m$ and satisfy the decoupled equations
\begin{equation}
i \frac{\partial \psi_n^{ ( \pm) }}{\partial m} = \pm \left( J_{n} \psi_{n+1}^{( \pm)}+J_{n-1} \psi_{n-1}^{(\pm)} \right) \label{S24}
\end{equation}
where we have set
\begin{equation}
J_n=\frac{1}{2} \theta_{n+1} \simeq \frac{1}{2} \cos \beta_{n+1} \label{S25}
\end{equation}
and treated $m$ as continuous variable.
Clearly, Eq.(\ref{S24}) corresponds to the single-particle Schr\"odinger equation in a one-dimensional lattice with inhomogeneous hopping rates $J_n$. The off-diagonal Aubry-Andr\'e model is thus retrieved by letting $\beta_n= \pi/2 -2A-2B \cos ( 2 \pi \alpha n)$, with $A, B \ll 1$.

\subsection{S6. Incoherent photonic quantum walk}
When random phases $\phi_n^{(m)}$ are applied at every discrete time step $m$, the discrete-time incoherent quantum walk is  described by following map
 for the light pulse intensities $X_n^{(m)}=\overline{|u_n^{(m)}|^2}$ and $Y_n^{(m)}=\overline{|v_{n+1}^{(m)}|^2}$ in the two fiber loops
 \begin{eqnarray}
X_n^{(m+1)} & = &  \cos^2 \beta_{n+1} X_{n+1}^{(m)}+ \sin^2 \beta_{n+1} Y_{n}^{(m)}  \label{S26} \\
Y_n^{(m+1)} & = & \sin^2 \beta_{n} X_{n}^{(m)}+ \cos^2 \beta_{n} Y_{n-1}^{(m)}  \label{S27}
\end{eqnarray}
 where the overline denotes statistical average. The above equations describe a classical random walk on two lattices and are analogous to Eq.(S14) for a binary lattice system. They conserve the particle probability, $\sum_n( X_n^{(m)}+Y_n^{(m)})=1$, and are readily obtained after taking the modulus square of both sides in Eqs.(\ref{S19}) and (\ref{S20}) and making the statistical average, using the property that $\overline{u_n^{(m)} v_n^{(m)*}}=0$. Let $X_n^{(m)}=\mu^m X_n $ and $Y_n^{(m)}=\mu^m Y_n$, where $\mu$ and $(X_n,Y_n)$ are the $2L$ eigenvalues and corresponding eigenvectors  of the incoherent propagator $\mathcal{U}$ defined by Eqs.(\ref{S26}) and (\ref{S27}) , i.e.
 \begin{equation}
  \mathcal{U}\left(
 \begin{array}{c}
 X_n \\
 Y_n
 \end{array}
 \right)=\mu \left(
 \begin{array}{c}
 X_n \\
 Y_n
 \end{array}
 \right). \label{S28}
 \end{equation}
  Owing to particle probability conservation, the eigenvalues $\mu$ satisfy the constraint $|\mu| \leq 1$, and there is always the eigenvalue $\mu_1=1$ with the corresponding eigenvector $X_n=Y_n=1/(2L)$, where $L$ is the number of lattice sites (periodic boundary conditions are assumed and $L$ is a Fibonacci number used to approximate the inverse of the golden ratio $\alpha=( \sqrt{5}-1)/2$).
  This is the most dominant (i.e. with the largest $|\mu|$) and non-decaying eigenstate of $\mathcal{U}$, corresponding to uniform excitation of the lattice.
  This means that asymptotically any initially localized excitation in the lattice spreads to reach a uniform distribution. However, if there are localized eigenstates of the transition matrix with extremely long lifetimes, i.e. with $| \mu|$ very close to 1, the spreading can be greatly slowed down and excitation can be transiently trapped in the lattice.\\
  Clearly, when the coupling angle $\beta_n$  is given by Eq.(\ref{S21}) with $| \theta _n| \sim \epsilon \ll 1$ and the photonic quantum walk emulates the off-diagonal Aubry-Andr\'e model, the incoherent dynamics described by the map (\ref{S26}) and (\ref{S27}) should correspond to the classical master equation given by Eqs.(4) and (5) of the main text. In fact, in this case we can eliminate from Eqs.(\ref{S26}) and (\ref{S27}) the variables $Y_n^{(m)}$, and up to order $\sim \epsilon$ one obtains the second-order difference equation for $X_n^{(m)}$
  \begin{equation}
  X_{n}^{(m+1)}=X_{n}^{(m-1)}+ \theta_{n+1}^2 X_{n+1}^{(m)}+\theta_n^2 X_{n-1}^{(m)}-(\theta_n^2+\theta_{n+1}^2) X_n^{(m-1)} \label{S29}
  \end{equation}
  with $Y_{n}^{(m+1)} \simeq X_{n}^{(m)}$. This means that, after each time step, the light pulse distribution in the two loops (sublattices), i.e. variables $X_n$ and $Y_n$, are flipped. A similar second-order difference equation is obtained for $Y_n^{(m)}$, namely
  \begin{equation}
  Y_{n}^{(m+1)}=Y_{n}^{(m-1)}+ \theta_{n+1}^2 Y_{n+1}^{(m)}+\theta_n^2 Y_{n-1}^{(m)}-(\theta_n^2+\theta_{n+1}^2) Y_n^{(m-1)}. \label{S30}
  \end{equation}
  If we sum both sides of Eqs.(\ref{S29}) and (\ref{S30}), one obtains the following difference equation
  \begin{equation}
  P_{n}^{(m+1)}=P_{n}^{(m-1)}+ \theta_{n+1}^2 P_{n+1}^{(m)}+\theta_n^2 P_{n-1}^{(m)}-(\theta_n^2+\theta_{n+1}^2) P_n^{(m-1)}. \label{S31}
  \end{equation}
  for the probabilities $P_n^{(m)}=X_n^{(m)}+Y_n^{(m)}$. Since $P_n^{(m)}$ varies slowly at each time step, we can treat $m=t$ as a continuous time variable, and after letting $P_n(t)=P_n^{(m)}$ the difference equation (\ref{S31}) can be approximated by the differential equation
  \begin{equation}
  \frac{dP_n}{dt}=-\frac{1}{2} ( \theta_n^2+\theta_{n+1}^2) P_n+ \frac{1}{2} \theta_{n+1}^{2} P_{n+1}+\frac{1}{2} \theta_n^2 P_n \label{S32}
\end{equation}  
which is precisely the classical master equation, defined by Eqs.(4) and (5) of the main text, with $\gamma=1$ and $J_n=\frac{1}{2} \theta_{n+1} \simeq \frac{1}{2} \cos \beta_{n+1}$. 

\end{widetext}

\end{document}